\documentclass[twocolumn]{aastex701}

\shorttitle{Interpreting Metallicity Gradient Residuals}
\shortauthors{Modak et al.}
\graphicspath{{./}{figures/}}

\usepackage{graphicx}
\usepackage{bm}
\usepackage{hyperref}
\usepackage{float}
\usepackage{amsmath, amssymb}
\usepackage{xcolor}
\usepackage{enumitem}

\usepackage[normalem]{ulem}

\newcommand{\md}{\mathrm{d}}

\newcommand{\br}{\bm{r}}
\newcommand{\bvee}{\bm{v}}

\newcommand{\nn}{\nonumber}
\newcommand{\FeH}{[\mathrm{Fe/H}]}
\newcommand{\ptot}{p_\mathrm{tot}}
\newcommand{\sigmatot}{\sigma_\mathrm{tot}}

\begin{document}

\title{An Analytic Model for Stellar Metallicity Gradient Residuals in Cold, Phase-Mixed Galactic Disks}

\author[0000-0002-8532-827X]{Shaunak Modak}
\affiliation{Department of Astrophysical Sciences, 4 Ivy Lane, Princeton University, Princeton, NJ 08544, USA}
\email{shaunakmodak@princeton.edu}

\author[0000-0002-3855-3060]{Zoe Hackshaw}
\affiliation{Department of Astronomy, The University of Texas at Austin, 2515 Speedway Boulevard, Austin, TX 78712, USA}
\email{zoehackshaw@utexas.edu}

\author[0000-0002-1423-2174]{Keith Hawkins}
\affiliation{Department of Astronomy, The University of Texas at Austin, 2515 Speedway Boulevard, Austin, TX 78712, USA}
\email{keithhawkins@utexas.edu}

\begin{abstract}

The distribution of stellar metallicities over phase space in galactic disks is sculpted by both star formation and secular orbital transport processes. As a result, chemo-dynamical models that infer radial heating and migration histories from observations typically rely on sophisticated numerical modeling of stellar distribution functions, star formation histories, and various dynamical perturbations. Here, we develop a complementary minimal analytic model for constraining radial migration, using stellar metallicity residuals relative to overarching galactic metallicity gradients. Incorporating residuals imprinted during formation and produced dynamically, and assuming a cold, phase-mixed stellar disk, we derive a closed-form expression for the resulting residual distribution. Applying our model to observed [Fe/H] residuals for stars in the Galactic thin disk, we find the root-mean-square amplitude of radial migration for stars of age $\tau$ to be $\langle (\delta R_\mathrm{g})^2 \rangle^{1/2} \approx (2.79 \pm 0.07)\,\mathrm{kpc} \times [\tau/(6\,\mathrm{Gyr})]^{1/2}$, consistent with results derived from more complex numerical frameworks. Our results clarify the physical origins of covariances and degeneracies common across chemo-dynamical transport models, and demonstrate that metallicity residuals provide a flexible, interpretable probe of radial migration in galactic disks.

\end{abstract}


\section{Introduction}
\label{sec:introduction}

The joint dynamical and chemical structure of a galaxy's constituent stellar populations encodes a wealth of information about its formation and evolution (see, e.g., the classic review by \citealt{FreemanBlandHawthorn2002}). Enabled by modern wide-field surveys like \textit{Gaia} \citep{Gaia} and APOGEE \citep{APOGEE}, ``chemo-dynamical'' studies of stellar abundance trends across phase space have yielded constraints on the density profile of the Milky Way's dark matter halo (e.g., \citealt{Horta2024, Horta2026}), the amplitude and morphology of its bar and spiral structure (e.g., \citealt{Chiba2021, Filion2023, Chiba2026, Jurado2026}), its ``inside-out'' formation (e.g., \citealt{Hayden2015, Bovy2016, Frankel2019}), and its recent interaction history (e.g., \citealt{Carr2022, Florent2025, Ding2025}).

Given measurements of stellar positions $\{\br_i\}$, velocities $\{\bvee_i\}$, ages $\{\tau_i\}$, and metallicities\footnote{In practice, it is common to use $\mu = \FeH$, which together with $\tau$ encodes essentially the entire chemo-dynamical state of stars in the Galaxy's low-$\alpha$ disk \citep{Ness2019}, but the analysis we present applies equally well to any abundance ratio, so we use the more general shorthand $\mu$ for ease of notation.} $\{\mu_i\}$, these inferences use each star's metallicity $\mu_i$ and age $\tau_i$ to to identify its co-natal population (e.g., \citealt{BlandHawthorn2010, Hogg2016}) and estimate its birth position (e.g., \citealt{Minchev2018, Lu2024}), and model how various dynamical processes have transported it to its observed phase-space position $(\br_i, \bvee_i)$, typically using $N$-body simulations (e.g., \citealt{DiMatteo2013, Minchev2013, Khoperskov2018, Debattista2025}). For a star in the Galactic disk plane, this transport may be decomposed into radial heating (changes in the extent of its deviations from a circular orbit) and radial migration (changes in its orbital angular momentum). In the angle-action framework, these correspond to changes in the radial and azimuthal actions $J_R$ and $J_\varphi$ respectively, which remain invariant in the absence of perturbations (e.g., \citealt{BinneyLacey1988, Sellwood2014}; C. Hamilton et al., in preparation). Since stars are born with some intrinsic metallicity scatter (e.g., \citealt{DeCia2021}) and experience different realizations of (stochastic) time-dependent perturbations as they orbit, the transport history of any individual star is generally difficult to reconstruct uniquely. Instead, chemo-dynamical studies typically characterize transport using population-level correlations in large stellar ensembles, where radial heating corresponds to a change in the ensemble's radial velocity dispersion, and radial migration corresponds to a change in the variance and/or mean of the ensemble's orbital angular momentum distribution.

However, interpreting these population-level trends is challenging, because the observed stellar distribution function (DF) is shaped not only by heating and migration, but also the galaxy's star formation history and chemical evolution. Efforts to constrain orbital transport in the Milky Way (e.g., \citealt{Frankel2018, Frankel2019, Frankel2020, Lian2022, Ratcliffe2025, Zhang2025}) have therefore relied on sophisticated numerical models that simultaneously fit all of these processes either with prescribed functional forms, or flexible but even more complex non-parametric frameworks. While these approaches have yielded many important insights, they typically require careful statistical inference techniques to handle their high-dimensional parameter spaces, and the complexity of the methods can sometimes obscure the relationship between the inferred parameters and the underlying physics.

One emerging approach to disentangling these processes is to recast the problem in terms of deviations from a smooth background metallicity profile across the galaxy (e.g., \citealt{Poggio2022, Hawkins2023, Hackshaw2024, Barbillon2025}). That is, from the positions $\{\br_i\}$ and metallicities $\{\mu_i\}$, one first constructs a model $\hat{\mu}(\br)$ that captures large-scale trends in metallicity, then studies the statistics of the \textit{residuals} $\{\delta\mu_i \equiv \mu_i - \hat{\mu}(\br_i)\}$. The model $\hat{\mu}$ is typically a simple function of the galactocentric radius $R$ and height from the midplane $|z|$ alone, encoding the Galaxy's dominant radial and vertical gradients (e.g., \citealt{Janes1979, Rolleston2000, Otto2026}). Consequently, the residuals are highly structured, and the system's dynamical history may be inferred from their features. For instance, $N$-body simulations have demonstrated that azimuthal metallicity-residual variations can arise from perturbations driven by the Galactic bar and spiral structure (e.g., \citealt{Debattista2025, Jurado2026}).

In this work, we develop a stripped-down, analytic approach for interpreting metallicity residuals to probe radial migration in galactic disks, complementary to existing numerical and statistical modeling efforts. Physically, we decompose a star's metallicity residual into (i) variations imprinted at formation, (ii) variations produced by radial migration, and (iii) variations resulting from its radial orbital motion, and develop a simple theoretical framework incorporating all three effects. Taking the stellar orbits to be nearly circular, and assuming the stellar population to be uniformly distributed in orbital phase, the resulting residual distribution takes a closed functional form. Thus, we provide an interpretable connection between observed metallicity residuals and radial migration in the disk.

The remainder of this Letter is organized as follows. In \autoref{sec:model}, we describe our underlying assumptions and approximations, apply them to derive an analytic model for the distribution of stellar metallicity residuals, and highlight the model's key features. In \autoref{sec:application}, we compare our model against observations of residuals in $\mu = \FeH$ for stars in the Milky Way's thin disk, and apply it to quantify radial migration in the disk. In \autoref{sec:summary_discussion}, we summarize, discuss limitations of the model, and outline possible extensions of our analysis.

\section{Modeling Metallicity Residuals}
\label{sec:model}

\subsection{Assumptions and approximations}
\label{sec:assumptions_approximations}

To render our model for metallicity residuals tractable, we make several well-motivated simplifying assumptions. First, we restrict our analysis to stars on kinematically cold orbits that do not make significant excursions from the Galactic mid-plane. Thus, we neglect the vertical dynamics of the problem, and use the epicycle approximation (e.g., \citealt{BT}) to describe the radial motion:
\begin{align}
    \label{eq:epicycle_R}
    R(t) & = R_\mathrm{g} - a_R\cos\theta_R(t), \\
    \label{eq:epicycle_theta}
    \theta_R(t) & = \kappa(R_\mathrm{g})t + \theta_0,
\end{align}
where $R_\mathrm{g}$ is the star's guiding radius, $a_R$ is its epicyclic amplitude, $\theta_R$ is the radial angle (with some initial orbital phase $\theta_0$), and
\begin{equation}
    \label{eq:kappa_definition}
    \kappa(R) \equiv \sqrt{2}\,\frac{v_c(R)}{R}\left(1 + \frac{\md\log v_c(R)}{\md\log R}\right)^{1/2}
\end{equation}
is the epicyclic frequency set by the galactic rotation curve $v_c(R)$. This approximation neglects corrections of $\mathcal{O}((a_R/R_\mathrm{g})^2)$ to the star's orbit. Epicyclic orbits are characterized by their azimuthal actions $J_\varphi \equiv R_\mathrm{g} v_c(R_\mathrm{g})$ and radial actions $J_R \equiv (1/2)\kappa(R_\mathrm{g}) a_R^2$. Although many studies focus on metallicity residual trends in action space (e.g., \citealt{Hackshaw2024, Debattista2025}), the model we develop depends on the actions only through $a_R \equiv \sqrt{2J_R/\kappa(J_\varphi)}$, so we work with $a_R$ directly.

In addition, we assume that the stellar population we study is radially phase-mixed, i.e., the stellar DF is independent of $\theta_R$. For reference, in a phase-mixed ensemble at a fixed guiding radius, the root-mean-square epicylic amplitude is related to the radial velocity dispersion by
\begin{equation}
    \label{eq:radial_velocity_dispersion}
    \langle a_R^2 \rangle^{1/2} = \sqrt{2}\frac{\sigma_R}{\kappa(R_\mathrm{g})},
\end{equation}
where here and for the remainder of this paper, the angle brackets denote an ensemble-average. By Jeans theorem, equilibrium stellar DFs satisfy this phase-mixed criterion. Of course, the Galaxy is \textit{not} entirely in equilibrium (e.g., \citealt{Antoja2018}), but we demonstrate in \autoref{sec:application} that this approximation still captures many observed trends.

Finally, because we are neglecting the vertical dynamics here, our metallicity model is a function $\hat{\mu}(R)$ of galactocentric radius alone. Following convention from the literature (e.g., \citealt{Janes1979, Rolleston2000, Hawkins2023, Hackshaw2024}), we use a linear model,
\begin{equation}
    \label{eq:linear_metallicity}
    \hat{\mu}(R) = \mu_0 + m_R R,
\end{equation}
where $m_R < 0$ is the metallicity gradient. In principle, $\mu_0$ and $m_R$ may vary over secular timescales, but our analysis will consider mono-age populations, and thus we suppress this time-dependence for ease of notation.

\subsection{Residuals imprinted during formation}
\label{sec:residual_formation}

We assume that the scatter in stellar birth metallicities is inherited from the metallicity scatter in the gas from which the stars form, and model this contribution to the metallicity residual, $\delta\mu_\mathrm{f}$, as drawn from a Gaussian with zero mean and a standard deviation $\sigma_\mathrm{f}$. The value of the scatter at a given radius is measured to depend only very weakly on radius in gas-phase metallicity observations of nearby galaxies (e.g., \citealt{Kreckel2019}) as well as in the FIRE simulations of Milky-Way-like systems (e.g., \citealt{Bellardini2021, Bellardini2022}), so we approximate $\sigma_\mathrm{f}$ to be independent of galactocentric radius. We also assume $\sigma_\mathrm{f}$ to be age-independent, but this approximation may be trivially relaxed, as we discuss in \autoref{sec:application}. Thus, the distribution of metallicity residuals imprinted by formation alone is 
\begin{equation}
    \label{eq:pdf_formation}
    p_\mathrm{f}(\delta\mu_\mathrm{f}) \equiv \frac{1}{\sqrt{2\pi \sigma_\mathrm{f}^2}} \exp\left(-\frac{\delta\mu_\mathrm{f}^2}{2\sigma_\mathrm{f}^2}\right).
\end{equation}

\subsection{Residuals produced dynamically}
\label{sec:residual_dynamics}

For the linear model given by \autoref{eq:linear_metallicity}, metallicity residuals of $m_R\delta R$ are produced dynamically when the present day observed galactocentric radius of a star differs by $\delta R$ from its birth radius. Using \autoref{eq:epicycle_R}, we can decompose the difference $\delta R$ into contributions from (i) radial migration, a change in the star's guiding radius $\delta R_\mathrm{g}$, and (ii) epicyclic blurring, a deviation of the star from its present-day guiding radius, $-a_R\cos\theta_R(t)$.

We model the outcome of radial migration using a Gaussian distribution with zero mean and age-dependent variance $\langle(\delta R_\mathrm{g})^2\rangle(\tau)$, which is a generic outcome of diffusive radial transport driven by local perturbations like the interstellar medium and quasilinear spiral structure, as well as impulsive velocity kicks arising from compact scatterers (\citealt{Modak2026radial}; C. Hamilton et al., in preparation). In particular, writing $\delta\mu_\mathrm{m} \equiv m_R\delta R_\mathrm{g}$, we have
\begin{equation}
    \label{eq:pdf_migration}
    p_\mathrm{m}(\delta\mu_\mathrm{m}; \tau) = \frac{1}{\sqrt{2\pi\sigma_\mathrm{m}(\tau)^2}}\exp\left(-\frac{\delta\mu_\mathrm{m}^2}{2\sigma_\mathrm{m}(\tau)^2}\right),
\end{equation}
where
\begin{equation}
    \label{eq:sigma_m_definition}
    \sigma_\mathrm{m}^2 = m_R^2 \langle(\delta R_\mathrm{g})^2\rangle.
\end{equation}
The age-dependence of $\sigma_\mathrm{m}$ depends on the microphysics of the perturbations driving migration: for instance, in a flat rotation curve, we expect a random-walk scaling $\langle(\delta R_\mathrm{g})^2\rangle \propto \tau$ for circular orbits, but a sub-diffusive scaling $\langle(\delta R_\mathrm{g})^2\rangle \propto \tau^{4/5}$ once the typical epicyclic amplitude of stars exceeds the spatial scale of the perturbation (\citealt{GK1, Modak2026radial}; C. Hamilton et al., in preparation). Note that we need not \textit{prescribe} the age-dependence: instead, we can constrain $\sigma_\mathrm{m}(\tau)$ by fitting residual distributions across multiple age bins.

We model the contribution of epicyclic blurring, $\delta\mu_\mathrm{e} \equiv -m_R a_R\cos\theta_R$ using the phase-mixed assumption: for $\theta_R$ uniformly distributed over $[0, 2\pi)$, we have
\begin{equation}
    \label{eq:pdf_epicycles_twovalued}
    p_\mathrm{e}(\delta\mu_\mathrm{e}; a_R) = \frac{1}{2\pi}\left(\left|\frac{\md\delta\mu_e}{\md\theta_R}\right|^{-1}_{\theta_1} + \left|\frac{\md\delta\mu_e}{\md\theta_R}\right|^{-1}_{\theta_2}\right),
\end{equation}
where $\theta_1$ and $\theta_2$ are the two solutions of $\delta\mu_\mathrm{e} = -m_R a_R \cos\theta_R$. Evaluating the derivative then gives
\begin{equation}
    \label{eq:pdf_epicycles}
    p_\mathrm{e}(\delta\mu_\mathrm{e}; a_R) = \frac{1}{\pi |m_R| a_R} \left[1 - \left(\frac{\delta\mu_\mathrm{e}}{m_R a_R}\right)^2\right]^{-1/2}
\end{equation}
when $|\delta\mu_\mathrm{e}| \leq |m_R| a_R$, and $p_\mathrm{e} = 0$ for all $|\delta\mu_\mathrm{e}| > |m_R| a_R$. Again, we need not prescribe the form of the stellar DF or radial heating history here.

\subsection{The total metallicity residual distribution}
\label{sec:derivation_properties}

We can now combine the results of \autoref{sec:residual_formation} and \autoref{sec:residual_dynamics} to study the total residual,
\begin{equation}
    \label{eq:residual_decomposition}
    \delta\mu \equiv \delta\mu_\mathrm{f} + \delta\mu_\mathrm{m} + \delta\mu_\mathrm{e}.
\end{equation}
Because the contributions $\delta\mu_\mathrm{f}$, $\delta\mu_\mathrm{m}$, and $\delta\mu_\mathrm{e}$ are independently distributed (being set by the star's formation, migration history, and instantaneous orbital phase, respectively), from their pdfs $p_\mathrm{f}$, $p_\mathrm{m}$, and $p_\mathrm{e}$, we can immediately calculate that $\langle \delta\mu \rangle = 0$, as expected by construction of the linear fit $\hat{\mu}$ to the background metallicity profile. It is also straightforward to calculate the second moment,
\begin{equation}
    \label{eq:total_residual_variance}
    \langle \delta\mu^2 \rangle = \sigmatot(\tau)^2 + \frac{1}{2} m_R^2 a_R^2,
\end{equation}
where we have defined the total variance produced by formation and migration to be
\begin{equation}
    \label{eq:sigma_tot}
    \sigmatot(\tau)^2 \equiv \sigma_\mathrm{f}^2 + \sigma_\mathrm{m}(\tau)^2.
\end{equation}
While \autoref{eq:total_residual_variance} already provides a useful relationship between a stellar population's dynamical state and metallicity variation, we can proceed further by combining the three component pdfs to derive the full distribution of the total residual, providing a complete description beyond its initial moments.

\begin{figure*}
    \centering
    \includegraphics[width=0.99\textwidth]{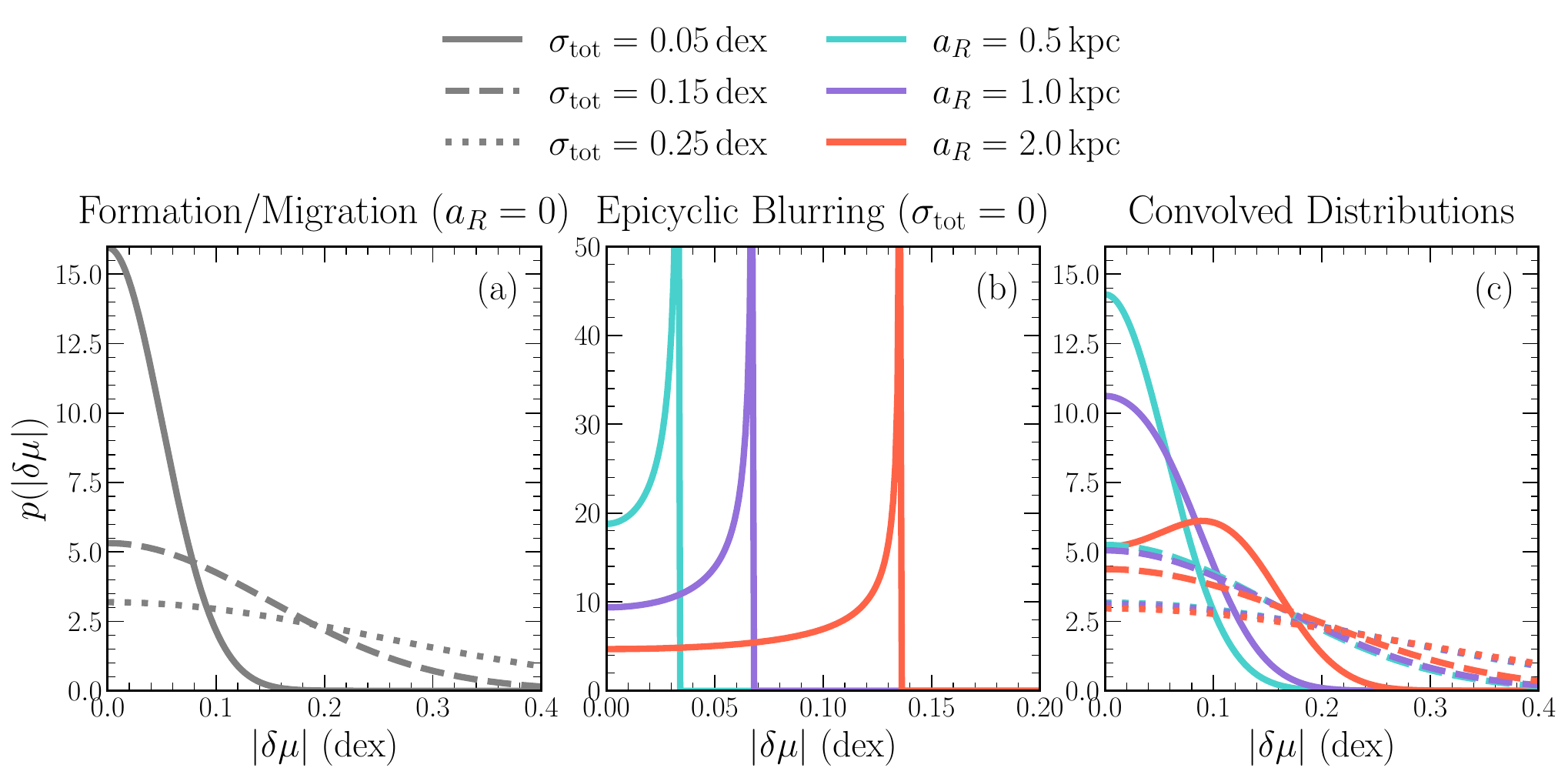}
    \caption{Absolute metallicity residual distributions produced by formation and migration processes (panel (a)), epicyclic blurring (panel (b)), and the combination of both effects (panel (c)). Different colors indicate different epicyclic amplitudes, while different line styles indicate different amounts of scatter from formation and migration. To produce these curves, we have used $m_R = -0.0678\,$dex/kpc as measured for $\mu = \FeH$ by \cite{Hackshaw2024}.}
    \label{fig:example_distributions}
\end{figure*}

Again using the fact that the contributions to the total residual are independent, the pdf of the summed residual is given by convolutions of the individual pdfs:
\begin{align}
    \label{eq:convolution}
    p_\mathrm{tot}(\delta\mu; a_R, & \tau) = \int \md(\delta\mu') p_\mathrm{e}(\delta\mu-\delta\mu'; a_R) \nn \\
    & \times \int \md (\delta\mu'') p_\mathrm{f}(\delta\mu'')p_\mathrm{m}(\delta\mu'-\delta\mu''; \tau).
\end{align}
The convolution of the two Gaussians $p_\mathrm{f}$ and $p_\mathrm{m}$ is simply another Gaussian with zero mean and variance $\sigmatot^2$. The convolution of the resulting Gaussian with $p_\mathrm{e}$ is more involved, so we defer the calculation to \autoref{sec:evaluating_convolution}, and just quote the result:
\begin{equation}
    \label{eq:pdf_total}
    p_\mathrm{tot}(|\delta\mu|; a_R, \tau) = 2 \times \frac{\chi(|\delta\mu|, a_R, \tau)}{\sqrt{2\pi\sigmatot(\tau)^2}} \exp\left(-\frac{|\delta\mu|^2}{2\sigmatot(\tau)^2}\right),
\end{equation}
where
\begin{align}
    \label{eq:chi_definition}
    \chi( & |\delta\mu|, a_R, \tau) \equiv \tilde{I}_0\left(\frac{1}{4}\left(\frac{m_R a_R}{\sigmatot(\tau)}\right)^2\right)I_0\left(\frac{m_R a_R|\delta\mu|}{\sigmatot(\tau)^2}\right) \nn \\
    & + 2\sum_{n=1}^{\infty}(-1)^n \tilde{I}_n\left(\frac{1}{4}\left(\frac{m_R a_R}{\sigmatot(\tau)}\right)^2\right)I_{2n}\left(\frac{m_R a_R|\delta\mu|}{\sigmatot(\tau)^2}\right).
\end{align}
Here, $I_n$ is the modified Bessel function of the first kind, and $\tilde{I}_n(x) \equiv e^{-x}I_n(x)$. Note that because the individual pdfs are all even functions, their convolution $p_\mathrm{tot}$ is also even. So, we follow the convention of \cite{Hackshaw2024} and focus on the \textit{absolute} residual $|\delta\mu|$ as indicated in the argument of $\ptot$ in \autoref{eq:pdf_total}, and have multiplied by two accordingly to properly normalize $\ptot$. \autoref{fig:example_distributions} illustrates the form of these distributions: panel (a) shows Gaussian distributions for several values of $\sigmatot$, panel (b) shows $p_\mathrm{e}$ for several values of $a_R$, and panel (c) shows the $p_\mathrm{tot}$ distributions produced by all combinations of those shown in panels (a) and (b).

The main advantage of this analytic calculation, complementary to more sophisticated models, is that it elucidates the underlying structure of the residual distributions and possible degeneracies that may arise when performing parameter inference. For example, in \autoref{eq:chi_definition}, we see that $m_R$ and $a_R$ only appear in the dimensionless combination
\begin{align}
    \label{eq:beta}
    \beta & \equiv \frac{m_R a_R}{\sigmatot},
\end{align}
so as long as $\sigmatot \gtrsim |m_R|a_R$, we expect stellar populations of different dynamical temperatures to be nearly indistinguishable by their residual distributions (see, e.g., the dashed and dotted curves in panel (c) of \autoref{fig:example_distributions}). Another degeneracy apparent from \autoref{eq:sigma_tot} is that for a mono-age stellar population, we cannot distinguish between residuals intrinsic to star formation and arising from migration, because $\sigma_\mathrm{f}$ and $\sigma_\mathrm{m}$ only appear in the combination $\sigmatot$. This degeneracy may be broken by studying stellar populations with different ages, as we discuss in \autoref{sec:application}. However, we see from \autoref{eq:sigma_m_definition} that there remains a degeneracy between the metallicity gradient and radial migration amplitude in determining $\sigma_\mathrm{m}$. Indeed, the covariance between those parameters is readily apparent in the posteriors of many chemo-dynamical fits (e.g., Figure 4 of \citealt{Frankel2020} or Figure 2 of \citealt{Zhang2025}).

Finally, with the expression for $\ptot$ given in \autoref{eq:pdf_total} in hand, it is possible to analytically evaluate further observables in addition to \autoref{eq:total_residual_variance}; as an example to enable a direct comparison with \cite{Hackshaw2024}, we calculate the mean absolute residual $\langle |\delta\mu| \rangle$ in \autoref{sec:exact_mean_absolute_residual}. The exact expression for $\langle |\delta\mu| \rangle$ is not particularly enlightening, so here we just highlight some useful limiting cases. For perfectly cold orbits (or perfectly flat metallicity gradients) where $\beta = 0$, we have $\chi = 1$, so $p_\mathrm{tot}$ is a folded Gaussian, and
\begin{equation}
    \label{eq:mean_residual_beta0}
    \langle|\delta\mu|\rangle = \sqrt{\frac{2}{\pi}}\sigmatot \quad\quad (\beta = 0).
\end{equation}
In the opposite limit $|\beta| \gg 1$, the distributions become peaked at a nonzero $|\delta\mu|$, and the distribution $p_\mathrm{tot}$ approaches $p_\mathrm{e}$, which leads to a mean absolute residual of
\begin{equation}
    \label{eq:mean_residual_betagg1}
    \langle|\delta\mu|\rangle \simeq \frac{2}{\pi}|m_R| a_R \quad\quad (|\beta| \gg 1).
\end{equation}
For the Galactic thin disk, this extreme is essentially never achieved: $|\beta|$ is typically small but nonzero, so a useful correction to \autoref{eq:mean_residual_beta0} with errors of $\mathcal{O}(\beta^4)$ is
\begin{equation}
    \label{eq:mean_residual_betall1}
    \langle|\delta\mu|\rangle \simeq \sqrt{\frac{2}{\pi}}\sigmatot \left[1 + \left(\frac{m_R a_R}{2\sigmatot}\right)^2\right] \quad\quad (|\beta| \ll 1).
\end{equation}
Comparing \autoref{eq:mean_residual_betagg1} and \autoref{eq:mean_residual_betall1}, we see that if the residuals are produced predominantly by blurring, we expect $\langle|\delta\mu|\rangle \propto a_R$ at all $a_R$, but if they instead arise mostly from formation and migration, we expect shallower increases in $\langle|\delta\mu|\rangle$ with $a_R$ at small $a_R$.

\section{Application to Milky Way Data}
\label{sec:application}

In this Section, we illustrate an application of the model developed in \autoref{sec:model} to interpret absolute residuals in $\mu = \FeH$ in the Milky Way, using a subset of the ``planar, thin-disk sample'' curated by \cite{Hackshaw2024} from APOGEE DR17 \citep{APOGEE_dr17}. Briefly, the sample derives abundances, ages, positions, and kinematics from the \texttt{astroNN} value-added catalogue \citep{LeungBovy2019}, and the transformation to angle-action space assumes the \texttt{MilkyWayPotential2022} mass model (fit to the rotation curve of \citealt{Eilers2019} and the local phase-spiral measurements of \citealt{DarraghFord2023}), as implemented in the \texttt{gala} code \citep{Gala}. Thin-disk stars are selected kinematically following \cite{Ramirez2013}, and the sample is restricted to stars with galactocentric radii $R \in (3.5, 15.5)\,\mathrm{kpc}$ and maximum excursion from the midplane $Z_\mathrm{max} \leq 0.3\,$kpc. Using this sample, \cite{Hackshaw2024} measured $m_R = -0.0678\pm 0.0004\,$dex/kpc for the $\FeH$ gradient. We adopt this value throughout the following analysis across all stellar ages for simplicity.

From this sample, we further restrict our consideration to stars with $a_R < 2\,$kpc to ensure that the epicyclic approximation is reasonable to apply---this final cut yields a sample of 23,137 stars ($\approx 91$\% of the full ``planar, thin-disk sample''), with $\langle a_R/R_\mathrm{g} \rangle \approx 0.1$. We subdivide this sample into twelve evenly spaced bins in $a_R$, as well as four age bins: $\tau < 2\,$Gyr, $\tau \in [2, 3.5)\,$Gyr, $\tau \in [3.5, 5)\,$Gyr, and $\tau > 5\,$Gyr. We adopt these broad age bins to reflect the typical $\sim 30\%$ uncertainties associated with these stellar age determinations \citep{Mackereth2019}. Note that these measurement uncertainties will cause some stars to scatter between bins, given that the bins are defined by measured rather than true ages. We discuss the potential impact of this contamination on our inferred age-dependence of $\sigmatot$ in \autoref{sec:summary_discussion}. With this binning, each of the 48 resulting $(a_R, \tau)$ bins contains at least 100 stars; the median number of stars per $(a_R, \tau)$ bin is 585. When fitting residual distributions, we represent the ensemble in each $(a_R, \tau)$ bin by the bin-center value of $a_R$ and the mean value $\langle\tau\rangle$ within that age bin.

\begin{figure}
    \centering
    \includegraphics[width=0.48\textwidth]{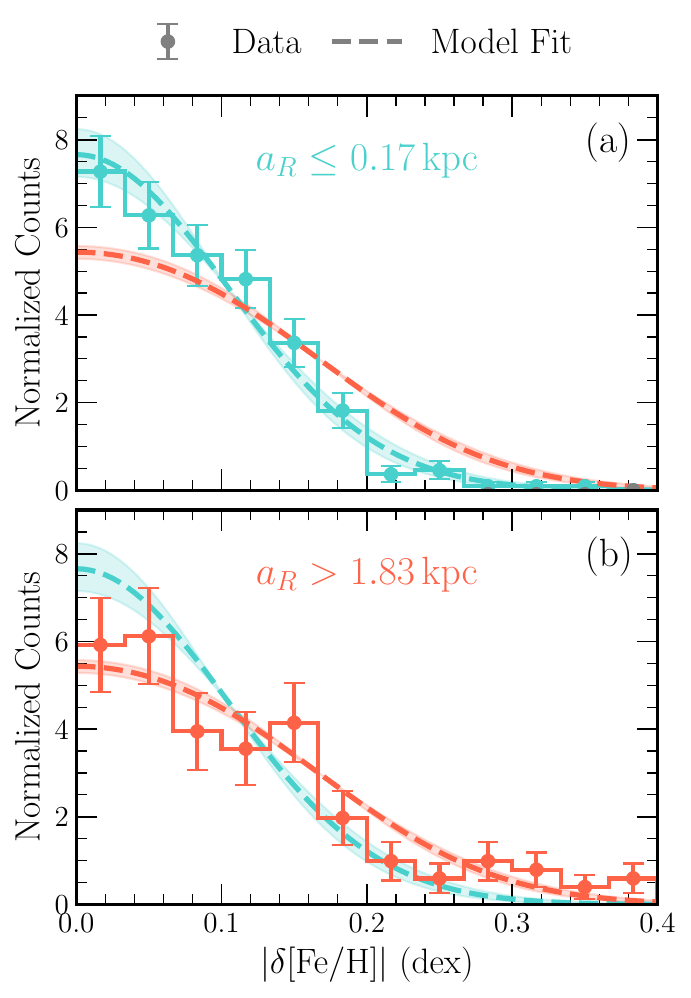}
    \caption{Absolute residual histograms (with poisson error bars) for stars with $\tau < 2\,$Gyr in the smallest (panel (a), cyan) and largest (panel (b), red) $a_R$ bins. The dashed curves show the model $p_\mathrm{tot}$ for each value of $a_R$ (in corresponding colors), using $m_R = -0.0678\,$dex/kpc and the fitted $\sigmatot$ value for this age bin. The shaded regions show the range of model distributions obtained by varying the fitted $\sigmatot$ by $\pm 1\sigma_{\sigmatot}$.}
    \label{fig:histogram_comparison}
\end{figure}

In \autoref{fig:histogram_comparison}, we show example absolute residual histograms for two ensembles of stars in the youngest age bin: those in the dynamically coldest (panel (a), cyan) and warmest (panel (b), red) $a_R$ bins. We see that the ensemble with larger $a_R$ has a broader residual distribution, as anticipated given the increased contribution of epicyclic blurring. Using analogous histograms, we fit a \textit{single} value of $\sigmatot$ to each of the four age bins by minimizing the weighted least-squares difference between the histograms and $\ptot$ across all twelve $a_R$ bins, with weights in each bin determined by Poisson uncertainties. We also estimate the statistical uncertainty $\sigma_{\sigmatot}$ on the fitted value of $\sigmatot$ in each bin using the parameter covariance matrix. The $p_\mathrm{tot}$ distributions using the best-fit value of $\sigmatot$ are shown in both panels as dashed curves in the same color as the histogram they are intended to fit for comparison; the shaded region around each curve indicates the variation in $p_\mathrm{tot}$ corresponding to the $\pm 1\sigma_{\sigmatot}$ range around the best-fit value. Especially in the high-$|\delta\FeH|$ tail, we see that the differences between the populations with different $a_R$ are reasonably well-captured by the model.

\begin{figure}
    \centering
    \includegraphics[width=0.48\textwidth]{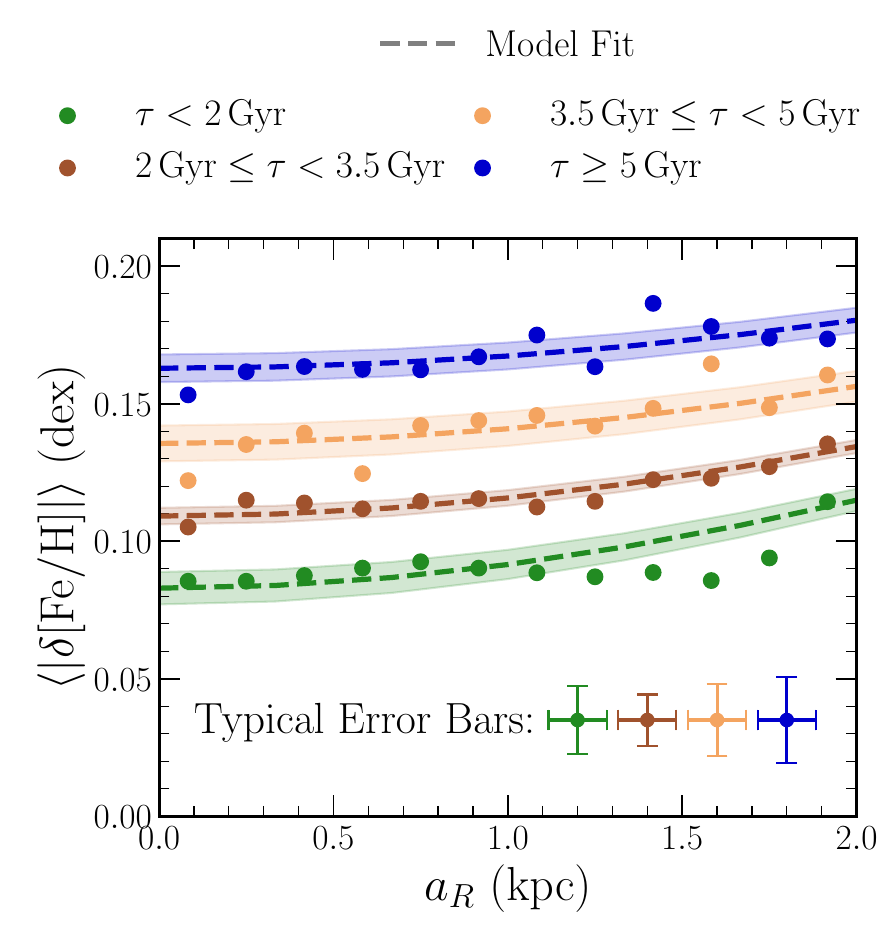}
    \caption{The mean absolute residuals in $\FeH$ as a function of $a_R$ for stars in different age bins (different colors); median errors on the mean for each age bin are shown at the bottom. The dashed curves show the predictions of \autoref{eq:mean_absolute_residual_series} with $m_R = -0.0678\,$dex/kpc and the best-fit value of $\sigmatot$ for each age bin; the shaded regions indicate the range of model curves obtained by varying the fitted $\sigmatot$ by $\pm 1\sigma_{\sigmatot}$.}
    \label{fig:mean_a_trend}
\end{figure}

In \autoref{fig:mean_a_trend}, we summarize the results of this fitting process for all $(a_R, \tau)$ bins by plotting the mean absolute residual in the bin as a function of $a_R$, with age bins indicated by different colors. For clarity, we do not plot error bars for each $(a_R, \tau)$ bin, and instead indicate the median error bar across all $a_R$ bins within a fixed $\tau$ bin at the bottom of the Figure. We see that older ensembles always exhibit significantly higher mean absolute residuals, as anticipated for metallicity variations driven by radial migration, which grow with $\tau$. For each age bin, the mean associated with the fitted pdf (calculated using \autoref{eq:mean_absolute_residual_series}) is shown as a function of $a_R$ with a dashed curve of the corresponding color. The shaded regions indicate the effect of propagating $\pm 1\sigma_{\sigmatot}$ errors on the values of $\sigmatot$ in each bin to $\langle|\delta\FeH|\rangle$. The model using the fitted $\sigmatot$ values captures the trend of increasing $\langle |\delta\FeH|\rangle$ with $a_R$ accurately for all stars with $\tau \gtrsim 2\,$Gyr, and stars with $\tau < 2\,$Gyr on relatively colder orbits ($a_R \lesssim 1\,$kpc). A slight departure (by $\lesssim 1.5\,\times\,$the errors on the means) from the model is evident for young stars on hotter orbits, which we attribute to a breakdown of the model's assumptions: to achieve such large epicyclic amplitudes at such young ages, these stars likely experienced strong, recent perturbations, and so they are likely not well phase-mixed.

\begin{figure}
    \centering
    \includegraphics[width=0.48\textwidth]{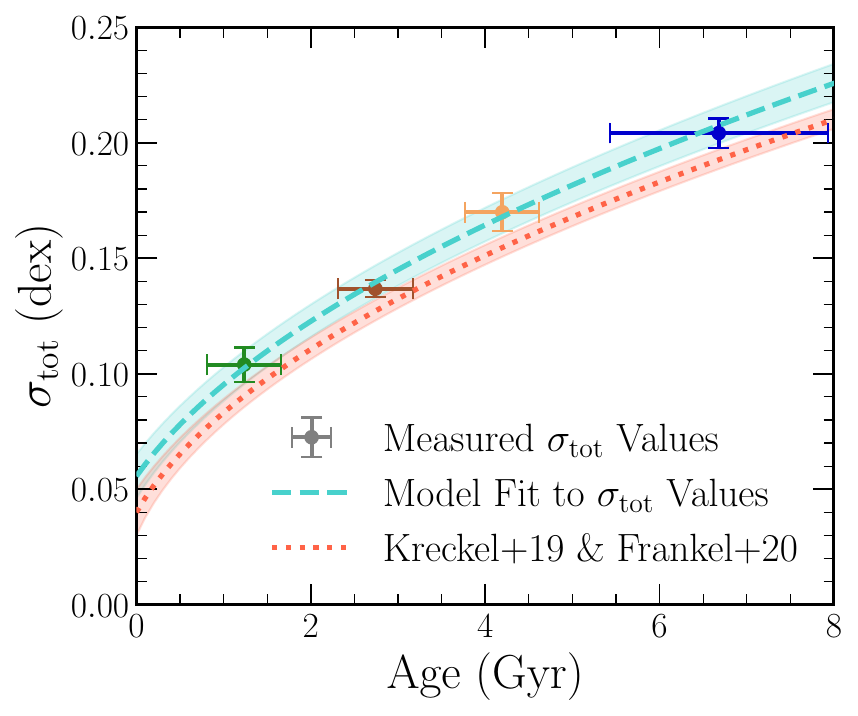}
    \caption{The measured values of $\sigmatot(\tau)$, with points colored as in \autoref{fig:mean_a_trend}. Horizontal error bars indicate the standard deviation of the ages in each bin, while vertical error bars indicate the $\pm 1\sigma$ range around the best-fit values of $\sigmatot$. The cyan dashed curve is the best-fit model of the form of \autoref{eq:sigma_tot_fitting}, while the red dotted curve shows the result of using parameters derived from \cite{Kreckel2019} and \cite{Frankel2020}. The shaded regions indicate the range of model curves obtained by varying the fitted $\sigma_{\mathrm{f}}$ and $\langle(\delta R_\mathrm{g})^2\rangle_0$ by their respective $\pm 1\sigma$ statistical uncertainties.}
    \label{fig:migration_measurement}
\end{figure}

Finally, in \autoref{fig:migration_measurement}, we plot the measured values of $\sigmatot$ as a function of $\langle\tau\rangle$ in each bin; the colors of each point correspond to those in \autoref{fig:mean_a_trend}. These values correspond to values of $|\beta|$ ranging from $\approx 0.03 - 1.25$, suggesting that formation and migration are almost always dominant over epicyclic blurring, except for the very youngest and dynamically warmest stars. Using \autoref{eq:sigma_m_definition} and \autoref{eq:sigma_tot}, we may interpret these results by fitting the points with a function of the form
\begin{equation}
    \label{eq:sigma_tot_fitting}
    \sigmatot(\tau) = \sqrt{\sigma_\mathrm{f}^2 + m_R^2 \langle(\delta R_\mathrm{g})^2\rangle_0 \left(\frac{\tau}{\tau_0}\right)^n},
\end{equation}
where $\langle(\delta R_\mathrm{g})^2\rangle_0$ and $\tau_0$ parametrize the amplitude of radial migration, and $n$ parametrizes its power law scaling. Of course, \autoref{eq:sigma_tot_fitting} is just one choice for modeling the variation of $\sigmatot$ in time: the measured $\sigmatot$ values may be more general functional forms corresponding to, e.g., age-dependent metallicity scatter at formation $\sigma_\mathrm{f}(\tau)$ or more varied radial migration statistics. However, to avoid overfitting our small number of age bins, we assume a constant $\sigma_\mathrm{f}$, and follow \cite{Frankel2020} in fixing $n = 1$ and $\tau_0 = 6\,$Gyr. With these choices, a weighted least-squares fit of the measured values of $\sigmatot$ (incorporating their uncertainties $\sigma_{\sigmatot}$, and the standard deviation of the ages within each age bin in the weighting) yields the best-fit parameters and $\pm 1\sigma$ statistical uncertainties
\begin{align}
    \label{eq:best_fit}
    \sigma_\mathrm{f} & \approx (0.056 \pm 0.009)\,\mathrm{dex}, \nn \\
    \langle(\delta R_\mathrm{g})^2\rangle^{1/2}_0 & \approx (2.79\pm 0.07)\,\mathrm{kpc};
\end{align}
the cyan dashed curve and shaded region in \autoref{fig:migration_measurement} shows this fit. The scatter at formation we measure is comparable\footnote{Note that we do not explicitly account for individual stellar abundance uncertainties in our residual-distribution fitting, so the fitted $\sigma_\mathrm{f}$ should be interpreted only as an upper limit on the true formation scatter.} to the values $\sigma_\mathrm{f}\sim 0.03-0.05\,$dex observed in nearby disks \citep{Kreckel2019} as well as the range of values $\sigma_\mathrm{f}\sim 0.04-0.08\,$dex produced in the FIRE simulations \citep{Bellardini2021, Bellardini2022}. Additionally, our best-fit migration amplitude is comparable to the values of $\langle(\delta R_\mathrm{g})^2\rangle^{1/2}_0 \approx (2.63\pm 0.03)\,$kpc found by \cite{Frankel2020} using \textit{Gaia} DR2 and APOGEE DR14, $\approx 2.6\,$kpc found by \cite{Ratcliffe2025} for stars with $\tau < 6\,$Gyr using APOGEE DR17, and $\approx 2.6\,$kpc for stars with guiding radii $R_\mathrm{g} \in [7.5, 8.5]\,$kpc found by \cite{Zhang2025} using \textit{Gaia} DR3 and LAMOST DR7.\footnote{\cite{Zhang2025} allowed for arbitrary age-dependence in $\langle(\delta R_\mathrm{g})^2\rangle(\tau)$, but for solar-neighborhood stars with ages from $2-8\,$Gyr, their results are reasonably well-described by a power law with index $n = 1$ (see their Figure 8), so we use that assumption to enable a comparison with our results.} For reference, the red dotted curve and shaded region in \autoref{fig:migration_measurement} shows the result of combining the \cite{Kreckel2019} and \cite{Frankel2020} results. We see that this combination is also reasonably consistent with our measured $\sigma_\mathrm{tot}$ points. Our measurements thus provide an independent assessment (using both a different dataset and different modeling technique) of the strength of radial migration.

\section{Summary and Discussion}
\label{sec:summary_discussion}

In this Letter, we have derived a simple model for the distribution of stellar metallicity residuals, incorporating both variations imprinted during star formation and those generated dynamically by radial migration and epicyclic blurring. In contrast to most other chemo-dynamical studies, our model is formulated directly in terms of epicyclic amplitudes and ages $(a_R, \tau)$, eliminating the need to specify stellar DFs or heating and migration histories \textit{a priori}, while remaining agnostic to the physical drivers of those processes. Because our model takes an analytic form, our results also provide insight into degeneracies in parameter estimation intrinsic to chemo-dynamical studies of radial transport. When applied to thin-disk stars on nearly-planar orbits, the model captures the broadening of residual distributions with increasing epicyclic amplitude, and yields an independent measurement of the amplitude of radial migration consistent with previous studies \citep{Frankel2020, Ratcliffe2025, Zhang2025}.

The key approximations enabling this analytic approach are (i) that stellar orbits are nearly planar and epicyclic, (ii) that the galactic stellar metallicity profile is described by a linear model with time-independent parameters, and (iii) that the stellar distribution is phase-mixed. We find that the epicycle approximation is applicable to $\sim 90\%$ of the thin-disk stars in the sample curated by \cite{Hackshaw2024}, and expect similar applicability to other thin-disk samples after appropriate $a_R$-cuts are applied. Our calculations can also be readily extended to vertical epicyclic motions, enabling an assessment of the contributions of vertical blurring to the scatter in elemental abundances that vary predominantly with $|z|$ rather than $R$, like [Mg/Fe] (see \citealt{Horta2026}). Additionally, our framework can be straightforwardly generalized beyond the constant-$m_R$ assumption (adopted in \autoref{sec:application} following the convention of \citealt{Frankel2020}) by fitting age-dependent metallicity gradients $m_R(\tau)$ and offsets $\mu_0(\tau)$ (see \autoref{eq:linear_metallicity}) before computing residuals to account for its evolution over time (e.g., \citealt{Minchev2018, Ratcliffe2023, Ratcliffe2025, Zhang2025}). We defer both this inclusion of an age-dependent metallicity profile and an extension of our calculations to 3D to future work.

With regard to phase-mixing, although we find trends in mean absolute residuals to be broadly well-described by our model (see \autoref{fig:mean_a_trend}), we are by construction unable to explain features like the extended ``clumps'' in residuals observed in, e.g., \cite{Hawkins2023, Hackshaw2024}. An analysis of these non-phase-mixed residuals is beyond the scope of this Letter (requiring either sophisticated numerical simulations as in \citealt{Debattista2025, Jurado2026}, or a linear perturbation theory building on the results presented here to incorporate non-uniform $\theta_R$ distributions in \autoref{eq:pdf_epicycles}), but we emphasize that our model provides a baseline distribution against which observations can be compared, enabling a cleaner isolation of non-equilibrium structure in the data.

Among the observational uncertainties limiting our approach, stellar age uncertainties are particularly dominant. These uncertainties effectively coarse-grain the recovered dependence of $\sigmatot$ on age, with the largest impact expected where the true underlying $\sigmatot(\tau)$ relation varies most rapidly. Our use of broad age bins mitigates, but does not eliminate this effect, so the inferred $\sigmatot(\tau)$ relation should be interpreted as a smoothed representation of the underlying age dependence. We expect age uncertainties to primarily modify the fine-grained age dependence rather than qualitatively change the overall increase of $\sigmatot$ with age that underlies our migration interpretation. A more precise treatment in future work would forward-model the age uncertainties and the underlying age distribution when fitting the metallicity-residual distributions.

Finally, our framework motivates two natural extensions to broader data sets and modeling applications. First, the formalism may be applied to external disk galaxies, even in the absence of full phase-space information. Integral field spectroscopic surveys such as SDSS-IV's MaNGA \citep{MaNGA} provide spatially resolved measurements of stellar velocity dispersions, rotation curves, and population-level abundances. Together, these observables enable estimates of representative epicyclic amplitudes in each spaxel via \autoref{eq:kappa_definition} and \autoref{eq:radial_velocity_dispersion}. A coarse observational proxy for the mean absolute metallicity residuals considered here would be the absolute residual between the mean stellar-population metallicity measured in each spaxel and a radial metallicity profile fitted across the galaxy, evaluated at the location of that spaxel. \autoref{eq:mean_residual_betall1} may then be used to estimate $\sigmatot$, which can be converted into an estimate of the age-averaged radial migration amplitude $\langle (\delta R_\mathrm{g})^2 \rangle^{1/2}$ using \autoref{eq:sigma_m_definition} and \autoref{eq:sigma_tot}, given an estimate for $\sigma_\mathrm{f}$ (e.g., \citealt{Kreckel2019}). The spaxel-level absolute residual is \textit{not} directly equivalent to the absolute mean of individual stellar residuals considered in our model, since each spaxel contains many unresolved stars and the measured metallicity is generally luminosity-weighted, but nevertheless, this observable may retain useful information about the spatial variation in metallicity residuals and could provide a first test of the model in constraining radial migration in external galaxies. A more rigorous approach could involve modeling the stellar populations within each spaxel and inferring a distribution of stellar metallicities rather than a single mean metallicity. Such distributions could be compared more directly with those predicted by our model, although this approach would be subject to additional degeneracies associated with the stellar-population modeling.

Second, for the Milky Way, where full phase-space information \textit{is} available, the distribution of metallicity residuals may be modeled to constrain $\sigmatot(\tau)$ jointly with the Galactic mass model and metallicity profile. Such an approach would incorporate stellar metallicities into dynamical models of the Galaxy in a physically motivated and largely analytically tractable framework, complementary to ``orbital torus imaging'' studies (e.g., \citealt{PriceWhelan2021, Horta2024, Horta2026}). Forthcoming data releases from \textit{Gaia} DR4 \citep{Gaia} and SDSS-V's Milky Way Mapper \citep{SDSSV, Ness2026b} will also substantially expand the sample of thin-disk stars, enabling finer resolution in both $a_R$ and $\tau$, and thereby a more detailed reconstruction of the Galaxy's migration history. This should in turn help elucidate the relative contributions of the Galactic bar (e.g., \citealt{Chiba2021, Filion2023, Chiba2026}), spiral structure (e.g., \citealt{SellwoodBinney2002, Roskar2012, VeraCiro2014, MeidtvanderWel2026}), and realistic ISM substructure (e.g., \citealt{Fujimoto2023, Modak2026radial, Modak2026characterizing}) to migration in the disk, without requiring assumptions about the stellar DF or the Galaxy's heating history. In this way, the analytic models for stellar metallicity residual distributions presented in this Letter offer a novel, broadly applicable probe of orbital transport in galactic disks. \newline

\begin{acknowledgments}
We thank Ivan Minchev, Adrian Price-Whelan, Carrie Filion, Chris Hamilton, Chris Carr, and Jenny Greene for helpful conversations, and the anonymous referee for a constructive report.
\end{acknowledgments}

\appendix

\section{Derivation of the Convolved pdf}
\label{sec:evaluating_convolution}

In this Appendix, we evaluate the convolution of the distribution $p_\mathrm{e}$ with a Gaussian distribution of metallicity residuals with mean $0$ and variance $\sigmatot^2$, which we denote $p_\mathrm{G}(\delta\mu_\mathrm{G}; \sigmatot^2)$. We have
\begin{align}
    p_\mathrm{tot}(\delta\mu) & = \int\md(\delta\mu') \, p_\mathrm{e}(\delta\mu') p_\mathrm{G}(\delta\mu' - \delta\mu; \sigmatot^2) \nn \\
    & = \int_{-\pi/2}^{\pi/2}\md u \, \frac{e^{-(m_R a_R \sin(u) - \delta\mu)^2/(2\sigmatot^2)}}{\pi\sqrt{2\pi\sigmatot^2}},
\end{align}
where in the second line we have plugged in the definitions of $p_\mathrm{e}$ and $p_\mathrm{G}$, and used the substitution $u = \arcsin(\delta\mu'/(m_R a_R))$. Expanding the argument of the exponent in the integrand, we find
\begin{equation}
    \label{eq:pre_Bessel_expansion}
    p_\mathrm{tot}(\delta\mu) = p_\mathrm{G}(\delta\mu; \sigmatot^2) \times \frac{e^{-\beta^2 / 4}}{\pi}\int_{-\pi/2}^{\pi/2}\md u \, e^{(\beta^2/4)\cos(2u) + (\beta\delta\mu/\sigmatot)\sin(u)},
\end{equation}
where we have defined $\beta$ as in \autoref{eq:beta}. The integral may now be evaluated by rewriting the integrand as a sum of modified Bessel functions of the first kind using the Jacobi-Anger expansion. Explicitly, using the relations
\begin{equation}
    e^{a\cos(2x)} = I_0(a) + 2\sum_{n=1}^{\infty}I_n(a)\cos(2nx)\mathrm{\quad\quad and\quad\quad} e^{b\sin(x)} = I_0(b) + 2\sum_{n=1}^{\infty}I_n(b)\cos(nx - n\pi/2)
\end{equation}
along with the orthogonality of the trigonometric functions, we can evaluate
\begin{equation}
    \frac{1}{\pi}\int_{-\pi/2}^{\pi/2} \md x \, e^{a\cos(2x) + b\sin(x)} = I_0(a) I_0(b) + 2\sum_{n=1}^{\infty}(-1)^n I_n(a) I_{2n}(b).
\end{equation}
Comparing with \autoref{eq:pre_Bessel_expansion}, we arrive at \autoref{eq:pdf_total}.

\section{Derivation of the mean absolute residual}
\label{sec:exact_mean_absolute_residual}

In this Appendix, we evaluate the mean absolute residual using the pdf $\ptot$ given in \autoref{eq:pdf_total}. Using the definition $\langle|\delta\mu|\rangle \equiv \int \md(|\delta\mu|) \, |\delta\mu| \ptot(|\delta\mu|)$, with the substitution $u = |\delta\mu|/\sigmatot$, we have
\begin{align}
    \label{eq:mean_absolute_residual_series}
    \langle|\delta\mu|\rangle & = \sqrt{\frac{2}{\pi}}\sigmatot \int_0^{\infty} \md u \, u\bigg[\tilde{I}_0\left(\frac{\beta^2}{4}\right)I_0(\beta u)e^{-u^2/2} + 2\sum_{n=1}^{\infty} (-1)^n \tilde{I}_n\left(\frac{\beta^2}{4}\right)I_{2n}(\beta u) e^{-u^2/2}
    \bigg] \nn \\
    & = \sqrt{\frac{2}{\pi}}\sigmatot \left[\tilde{I}_0\left(\frac{\beta^2}{4}\right) f_0 + 2\sum_{n=1}^{\infty}(-1)^n\tilde{I}_n\left(\frac{\beta^2}{4}\right) f_{2n}\right],
\end{align}
where
\begin{equation}
    f_n \equiv \int_0^\infty \md x \, xI_n(\beta x) e^{-x^2/2}.
\end{equation}
Using Equation 6.631 of \cite{GradshteynRyzhik}, we can evaluate
\begin{equation}
    f_{2n} = \left(\frac{\beta^2}{2}\right)^n \frac{\Gamma(1+n)}{\Gamma(1 + 2n)} e^{\beta^2/2} {}_1F_1\left(n; 2n+1; -\frac{\beta^2}{2}\right),
\end{equation}
where ${}_1F_1(a; b; z)$ is a confluent hypergeometric function. Note that the $n = 0$ term takes a particularly simple form, $f_0 = e^{\beta^2/2}$, while the contributions for $n \geq 1$ contain additional factors of $\beta^2$. So, in the limit $|\beta| \ll 1$ where $I_n(\beta^2/4) \propto (\beta^2/4)^n$, only the $n = 0$ term is relevant up to corrections of $\mathcal{O}(\beta^4)$, and we arrive at \autoref{eq:mean_residual_betall1} after further approximating $e^{\beta^2/4} \approx 1 + \beta^2/4$.

\bibliography{bibliography}{}
\bibliographystyle{aasjournalv7}

\end{document}